\documentclass[twocolumn,aps,prl,showpacs]{revtex4-2} 

\usepackage{graphicx,amsmath,amssymb,amsfonts,amsthm,xspace,soul}
\usepackage{dsfont,mathtools,xcolor,comment} 

\newtheorem{defn}{Definition}
\newtheorem{thm}{Theorem}
\newtheorem{lm}{Lemma}

\newcommand{\inner}[2]{\left\langle #1, #2 \right\rangle}
\newcommand{\ketbra}[2]{| #1\rangle \langle #2|}
\newcommand{\ket}[1]{| #1 \rangle}
\newcommand{\bra}[1]{\langle #1 |}
\newcommand{\kb}[1]{\ket{#1} \bra{#1}}
\newcommand{\id}{\mathds{1}} 
\newcommand{\prob}{\mathcal{P}}

\newcommand{\tr}{\mathrm{Tr}} 
\newcommand{\best}{\mathrm{Best}}
\newcommand{\worst}{\mathrm{Worst}}
\newcommand{\Pbest}{\prob_{\best}} 
\newcommand{\Pworst}{\prob_{\worst}} 
\newcommand{\PGM}{\mathrm{PGM}} 
\newcommand{\PBM}{\mathrm{PBM}}
\newcommand{\PPGM}{\prob_{\PGM}} 
\newcommand{\PPBM}{\prob_{\PBM}} 
 
\newcommand{\wt}[1]{\widetilde{#1}}

\begin{document} 

\title{The pretty bad measurement}

\author{Caleb McIrvin} 
\email{calebmcirvin111@vt.edu \vspace{0.25cm}}    
\author{Ankith Mohan}
\author{Jamie Sikora}

\affiliation{\vspace{0.25cm}Department of Computer Science, Virginia Tech, USA 24061 \vspace{0.25cm}}

\date{October 10, 2024} 

\begin{abstract} 
The quantum state discrimination problem has Alice sending a quantum state to Bob who wins if he correctly identifies the state. 
The pretty good measurement, also known as the square root measurement, performs pretty well at this task. 
We study the version of this problem where Bob tries to lose with the greatest probability possible (which is harder than it sounds). 
We define the pretty bad measurement which performs pretty well at this task, or in other words, pretty poorly for the original task. 
We show that both the pretty good measurement and the pretty bad measurement are always no worse than blind guessing at their respective tasks.  
As an application, we apply the pretty bad measurement to the quantum state anomaly detection problem and show how to avoid pretty bad qubits.\footnote{Published under the title ``Quantum state exclusion through offset measurement'' in \emph{Physical Review A}.} 
\end{abstract} 

\maketitle 

Preparing and measuring quantum states are the two pillars of quantum computing. 
As an example, quantum state tomography is the task of identifying which quantum state a device is preparing given (many) samples~\cite{christandl2012reliable, torlai2018neural, gross2010quantum, qi2013quantum, ahmed2021quantum, mahler2013adaptive, schwemmer2015systematic}
which has been experimentally tested~\cite{cramer2010efficient, stricker2022experimental, rambach2021robust, rippe2008experimental, mallet2011quantum}. 
In other words, one wishes to find the best measurement(s) to perform in order to learn the state with the fewest number of samples. 

Another task central to quantum computing is the \emph{quantum state discrimination problem} where one is tasked with identifying which state they have been given \emph{from a fixed set of quantum states}. 
This is a naturally occurring setting which has been studied in the context of quantum algorithms~\cite{radhakrishnan2009random, hayashi2006quantum}, quantum cryptography~\cite{van2002unambiguous, cabello2000quantum, ambainis2001new, sikora2017simple}, quantum machine learning~\cite{sentis2016quantum, sentis2017exact}, among many others (see~\cite{Bae_2015} for a survey).  

The quantum state discrimination problem can be phrased as a game between Alice and Bob. 
Alice has a finite set of quantum states $\{ \rho_1, \ldots, \rho_k \}$ and a fixed probability distribution $(p_1, \ldots, p_k)$, both of which are known to Bob. 
Alice then samples $i$ with probability $p_i$ and prepares and sends $\rho_i$ to Bob. 
Bob wins the game if he correctly identifies the state by sending the correct index, $i$, back to Alice. 
This game is illustrated below.  
\begin{figure}[h] 
    \centering
    \includegraphics[width=0.45\textwidth]{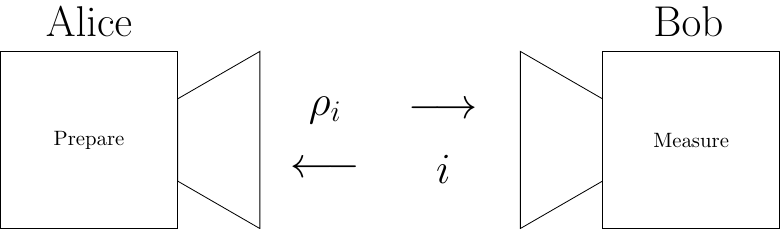}
    \caption{Alice prepares $\rho_i \in \{ \rho_1, \ldots, \rho_k \}$ and sends it to Bob. 
    Bob wins the game if he correctly identifies the index $i$.}
    \label{fig1}
\end{figure}

One can easily show that Bob can win this game perfectly if and only if the states have orthogonal supports, i.e., $\tr(\rho_i \rho_j) = 0$ for all $i \neq j$. 
When this is not the case, Bob may wish to maximize his probability of winning. 
Suppose he measures with the POVM $(M_1, \ldots, M_k)$ and on outcome $j$ he guesses the state was $\rho_j$. 
Then, given state $\rho_i$, he guesses ``I think the state is $\rho_j$'' with probability 
$\tr(\rho_i M_j)$.  
In this case, he wins the game with probability 
$\sum_{i=1}^k p_i \tr(\rho_i M_i)$.  
If Bob wishes to maximize this probability, then he is tasked with finding the best POVM $(M_1, \dots, M_k)$, known as the \emph{minimum error} measurement, and we define the best probability for Bob as 
\begin{equation}
    \label{eq:P_best} 
    \Pbest = \text{max} \; \sum_{i=1}^k p_i \tr(M_i \rho_i).
\end{equation} 

Given a problem instance, one can numerically solve for the best POVM using semidefinite programming~\cite{cvx}. 
Unfortunately, characterizing such POVMs is a difficult task known only to be possible in special cases, e.g., for $k=2$ the case of two states~\cite{helstrom1969quantum}, for $k$ arbitrary qubit states given with equal \emph{a priori} probabilities~\cite{deconinck2010qubit}, for geometrically uniform states~\cite{eldar2004optimal}, and mirror symmetric states~\cite{andersson2002minimum}.  
Therefore, it is important to find generic classes of measurements which have a simple form and \emph{approximate} $\Pbest$. 

\begin{figure}[h] 
    \centering
    \includegraphics[width=0.45\textwidth]{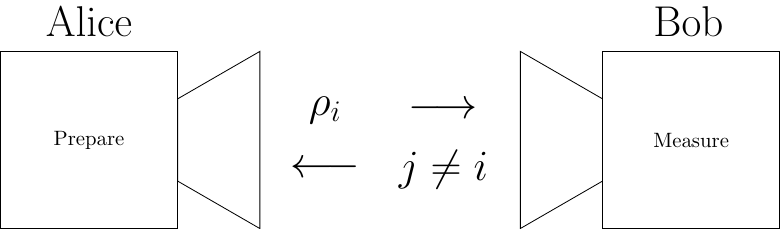}
    \caption{Alice prepares $\rho_i \in \{ \rho_1, \ldots, \rho_k \}$ and sends it to Bob. 
    Bob wins the game if he correctly \emph{excludes} an index $j \neq i$.}
    \label{fig2}
\end{figure}

\paragraph{Quantum state exclusion.} 
The quantum state \emph{exclusion} problem is similar to quantum state discrimination except that Bob wishes to guess a state \emph{which was not sent}. 
For example, if Alice sends $\rho_1$ to Bob and Bob responds with ``Alice, you did not send $\rho_3$'' then he would win the game in that case. 
The work in \cite{bandyopadhyay2014conclusive} studies this task in detail and its relationship to the PBR game~\cite{pusey2012reality}. 
Quantum state exclusion is illustrated in FIG.~\ref{fig2}.  

Unlike quantum state discrimination, it is much more difficult to determine when Bob has a perfect strategy to perfectly win this game, i.e., when he can exclude a state without error. 
If this is the case, we say that the states are \emph{antidistinguishable} which has been the topic of recent research~\cite{havlivcek2020simple, russo2023inner, leifer2020noncontextuality, mishra2023optimal, uola2020all, johnston2023tight}. 

If Bob wishes to find the measurement which best excludes a state, he wishes to minimize the probability of making an incorrect guess.
More precisely, if Bob measures $\rho_i$ with the POVM $(M_1, \ldots, M_k)$, he makes an \emph{incorrect guess} with probability $\tr(\rho_i M_i)$.  
Thus, he wishes to \emph{minimize} 
$\sum_{i=1}^k p_i \tr(\rho_i M_i)$.  

Notice that this is the exact opposite of what he wishes to do in the quantum state discrimination problem where he wants to maximize this quantity, see Eq.~\eqref{eq:P_best}. 
Thus, if Bob wants to exclude a state, his best measurement is also the \emph{worst} measurement for discrimination. 
Therefore, we define 
\begin{equation}
    \label{eq:P_worst}
    \Pworst = \text{min} \; \sum_{i=1}^k p_i \tr(M_i \rho_i). 
\end{equation}

It is worth noting that the viewpoint of Bob trying to lose the quantum state discrimination game is interesting in its own right. 
Indeed, Bob cannot always lose this game with certainty, even if he wants to, and understanding when this is the case, and more generally, understanding the above quantity gives insights into the very nature of quantum information.

The pretty good measurement (PGM), first discussed in~\cite{belavkin1975optimaldist, belavkin1975optimal, hughston1993complete}, is a way to approximate the best possible measurement. 
Sometimes PGM is optimal, some nice examples being for the hidden subgroup problem~\cite{bacon2005optimal, moore2005distinguishing}, for the quantum change point problem~\cite{sentis2016quantum}, and in port-based teleportation~\cite{leditzky2022optimality}.  
Generally speaking, for problems with a large amount of symmetry, the PGM seems to perform well. 

This brings us to the main question of this work:
\begin{flushleft}
\textit{Is there an analogue of the pretty good measurement for approximating $\Pworst$?} 
\end{flushleft}

We now review PGMs which helps us give an answer to the above question. 

\medskip 
\paragraph{The pretty good measurement (PGM).} 
Also known as the square root measurement, the pretty good measurement is the POVM $(G_1, \dots, G_k)$, where each measurement operator is defined as  
\begin{equation}
    \label{eq:PGM_ops}
    G_i = P^{-1/2} \bigg( p_i \rho_i \bigg) P^{-1/2} \; \text{ where } \; 
    P = \sum_{i=1}^k p_i \rho_i.
\end{equation} 
If Bob uses the PGM for quantum state discrimination, he wins with probability 
\begin{align} 
\PPGM 
& = \sum_{i=1}^k p_i \tr(\rho_i G_i) \\ 
& = \sum_{i=1}^k p_i^2 \tr(\rho_i P^{-1/2} \rho_i P^{-1/2}). \label{eq:P_PGM} 
\end{align} 
We now present a theorem which compares $\PPGM$ to the blind guessing probability, i.e., $1/k$. 

\begin{thm}{\cite{renes2017better}} \label{Thm:PGM}
For any set of states $\{ \rho_1, \ldots, \rho_k \}$ and probability distribution $(p_1, \ldots, p_k)$, we have 
\begin{equation} 
\PPGM \geq \frac{1}{k} + \frac{(1 - k \Pbest)^2}{k (k - 1)} 
\end{equation}
where $1/k$ can be interpreted as the blind guessing probability. 
Moreover, if $\PPGM = 1/k$, then we have $\rho_i = \rho_j$ and $p_i = p_j = 1/k$ for all $i \neq j$. 
\end{thm} 

We prove a slightly weaker version of this theorem in the appendix using the concept of \emph{offsets}, discussed shortly. 

We now discuss our first step towards defining a pretty bad measurement. 

\paragraph{Purposely messing up measurements with offsets.} 
Suppose that one is given a decent measurement for state discrimination and one wishes to make it perform worse. 
Then one way to do this is to take the outcome of this measurement and \emph{change the answer}. 
To this end, suppose we start with a POVM $(M_1, \ldots, M_k)$ and an \emph{offset parameter} $s$. 
We define the offset measurement $(M^s_1, \ldots, M^s_k)$ as 
\begin{equation} 
M^s_i = M_{i \oplus s} 
\end{equation} 
where we use the notation $\oplus$ to denote cyclic addition, e.g., $5 \oplus 1 = 1$ when $k = 5$. 
Conceptually, an offset should mess up a good measurement and, conversely, could have the ability to improve a bad measurement. 
We have the need to quantify how well these offset measurements perform, thus we define 
\begin{equation} 
\alpha_s = \sum_{i=1}^k p_i \tr(\rho_i M^s_i) = \sum_{i=1}^k p_i \tr(\rho_i M_{i \oplus s}).  
\end{equation} 
It turns out that all of these quantities are related to each other via the equality 
\begin{equation}\label{EqA3} 
\sum_{s=0}^{k-1} \alpha_s = 1 
\end{equation} 
which is easily seen by noting that, for any fixed $i$, $(M_i, M_{i \oplus 1}, \ldots, M_{i \oplus (k-1)})$ yields a valid POVM, and thus $\sum_{s=0}^{k-1} M_{i \oplus s} = \id$.

\paragraph{The pretty bad measurement (PBM).} 
We now define the counterpart to the PGM which now approximates $\Pworst$ instead. 
Our approach is quite simple and borrows much of the elegance from the PGM. 
We define it below and then give some interpretations of it. 
\begin{defn} 
Let $(G_1, \ldots, G_k)$ be the pretty good measurement. 
The pretty bad measurement (PBM) is defined as the POVM $(B_1, \ldots, B_k)$ where 
\begin{equation} 
B_i = \frac{1}{k-1} \bigg( \id - G_i \bigg).  
\end{equation} 
\end{defn} 

One way to interpret the PBM is by viewing it as the PGM with a uniformly random offset. 
In other words, 
\begin{align} 
B_i & = \frac{1}{k-1} \sum_{s=1}^{k-1} G^s_i 
\end{align} 
noting that we omit the $s=0$ index since $G_i^0 = G_i$.  
With this, we can calculate the success probability of the PBM as 
\begin{align} 
\PPBM 
& = \sum_{i=1}^k p_i \, \tr(\rho_i B_i) \\ 
& = \frac{1}{k-1} \sum_{s=1}^{k-1} \sum_{i=1}^k p_i \tr(\rho_i G^s_i) \\ 
& = \frac{1}{k-1} \sum_{s=1}^{k-1} \alpha_s. \label{JohnWick}
\end{align} 
Since $\alpha_0 = \PPGM$, as this is the no-offset success probability, we have the following equation which follows directly from Eq.~\eqref{EqA3}  
\begin{equation}
\PPGM + (k-1) \PPBM = 1. \label{PGM_PBM} 
\end{equation} 

By combining with Theorem~\ref{Thm:PGM}, we have our main result, below, which is the counterpart to the PGM bound, Theorem~\ref{Thm:PGM}. 

\begin{thm} \label{Thm:PBM}
For any set of states $\{ \rho_1, \ldots, \rho_k \}$ and probability distribution $(p_1, \ldots, p_k)$, we have 
\begin{equation} 
\PPBM \leq \frac{1}{k} - \frac{(1 - k \Pbest)^2}{k (k - 1)^2}. 
\end{equation} 
Moreover, if $\PPBM = 1/k$, then we have $\rho_i = \rho_j$ and $p_i = p_j = 1/k$ for all $i \neq j$. 
\end{thm} 

As a consequence of our discussion so far, we have 
\begin{equation}
\Pbest \geq \PPGM \geq \frac{1}{k} \geq \PPBM \geq \Pworst.
\label{AllInequalities}
\end{equation} 

\paragraph{Another interpretation.}  
There is another way to interpret the PBM which is as the PGM \emph{in a different setting}.  
For the set of states 
$\{ \rho_1, \ldots, \rho_k \}$ with probabilities 
$(p_1, \ldots, p_k)$, define a new problem instance 
$(\wt{\rho}_1, \ldots, \wt{\rho}_k)$ with probabilities 
$(\wt{p}_1, \ldots, \wt{p}_k)$ 
where 
\begin{equation} 
\wt{\rho}_i 
= \frac{1}{1-p_i} \sum_{j \neq i} p_j \rho_j 
\quad \text{ and } \quad
\wt{p}_i 
= \frac{1-p_i}{k-1} 
\end{equation} 
assuming $p_i \neq 1$ for all $i$. 
We can think of $\wt{\rho}_i$ as averaging over all of the $\rho_j$s with different indices. 
Thus, by guessing $i$ accurately in the tilde setting one is actually guessing poorly in the original setting.  
Before analyzing the PGM in the tilde setting, notice that 
\begin{equation}
\wt{p}_i \wt{\rho}_i 
= \frac{1}{k-1} \sum_{j \neq i} p_j \rho_j 
= \frac{1}{k-1} (P - p_i \rho_i)  
\end{equation}
recalling the definition of $P$ from Eq.~\eqref{eq:PGM_ops}. 
We also have that 
$\wt{P} 
= \sum_{i=1}^k \wt{p}_i \wt{\rho}_i 
= P$.   
Denoting the PGM in the tilde setting as $(\wt{G}_1, \ldots, \wt{G}_k)$, we can calculate that   
$\wt{G}_i 
= \wt{P}^{-1/2} \bigg( \wt{p}_i \wt{\rho}_i \bigg) \wt{P}^{-1/2}   
= B_i$. 
Thus, one can view the PBM as the PGM if the problem instance is changed accordingly.  

\smallskip 
\textit{Remark:} We note that the mixing of POVM elements considered above is studied more generally in the context of \emph{reverse channels} and \emph{reverse observables}
which have been used in the study of compatibility~\cite{filippov2017necessary}.  

\begin{figure}[h]
    \centering
    \includegraphics[width=0.45\textwidth]{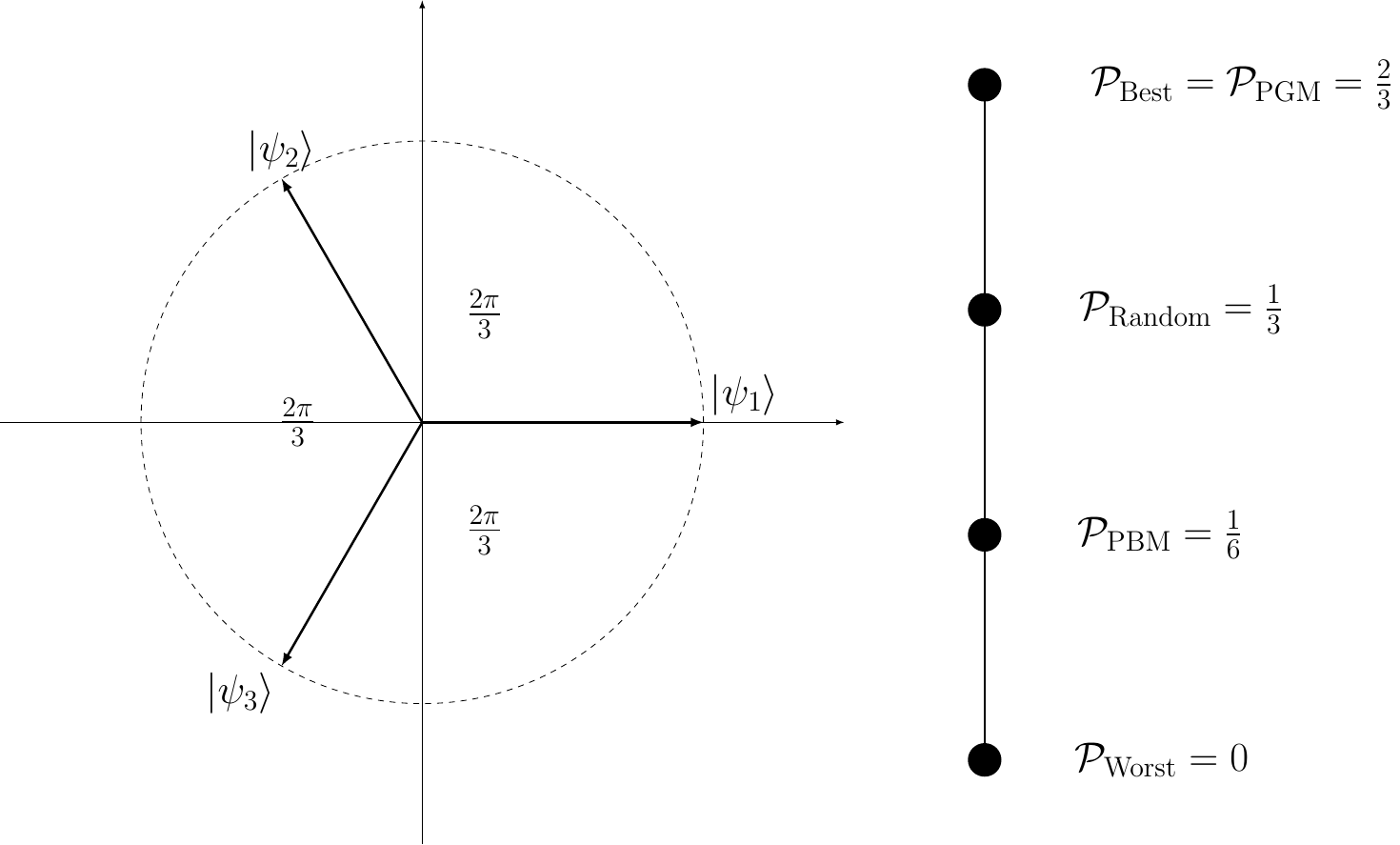}
    \caption{If given a trine state, each with equal probability, we have $\Pbest = \PPGM = 2/3$, 
    $\mathcal{P}_{\text{Random}} = 1/3$ being the blind guessing probability, and $\PPBM = 1/6$ approximating $\Pworst = 0$. 
    The pretty good measurement is given by the POVM $G_i = \frac{2}{3} \kb{\psi_i}$,
    the pretty bad measurement is given by the POVM $B_i = \frac{1}{3} \sum_{j \neq i} \kb{\psi_j}$,
    and the worst measurement is given by $W_i = \frac{2}{3} \kb{\psi_i^{\perp}}$, where $\kb{\psi_i^{\perp}}$ is a $\pi/2$-rotation of $\ket{\psi_i}$.}
    \label{fig:trine}
\end{figure}

\medskip 
\paragraph{Illustrative examples.}  
Consider the three trine states, defined as 
\begin{align} 
\ket{\psi_1} & = \ket{0}, \\
\ket{\psi_2} & = \cos(2\pi/3) \ket{0} + \sin(2\pi/3) \ket{1}, \\
\ket{\psi_3} & = \cos(4\pi/3) \ket{0} + \sin(4\pi/3) \ket{1}. 
\end{align} 
These are shown in FIG.~\ref{fig:trine}, along with the five quantities of interest.

We remark that this example sees the PGM being a perfect approximation of $\Pbest$ while the PBM is a decent, but not perfect, approximation of $\Pworst$. 
It is an interesting question to see in which cases the PBM is the worst measurement. 
The above example suggests that this is indeed tricky to answer since it does not coincide with the settings where the PGM is the best measurement. 
 
If the set of states is antidistinguishable, that is, if ${\Pworst = 0}$, then if $\PPBM = \Pworst$, then by Eq.~\eqref{PGM_PBM}, we must have $\PPGM = 1$.  
In this case, the states must have orthogonal supports and we have 
\begin{equation} 
\Pbest = \PPGM = 1 \; \text{ and } \; \PPBM = \Pworst = 0. 
\end{equation} 

On the other hand, if the set of states is maximally indistinguishable, i.e, $\rho_i = \rho_j$ and $p_i = p_j = 1/k$, for all $i \neq j$, then we have 
\begin{equation} \label{allthesame}
    \Pbest = \PPGM = \frac{1}{k} = \PPBM = \Pworst. 
\end{equation} 
What is interesting in the above two examples is that PBM being the worst measurement implies that PGM is the best measurement. 
This is actually true in general, which we state below and prove in the appendix, however the trine state example shows that the converse is not true. 

\begin{lm}
If $\PPBM = \Pworst$, then $\PPGM = \Pbest$.  
\end{lm} 

\begin{figure}[h]
    \centering
\includegraphics[scale=0.220]{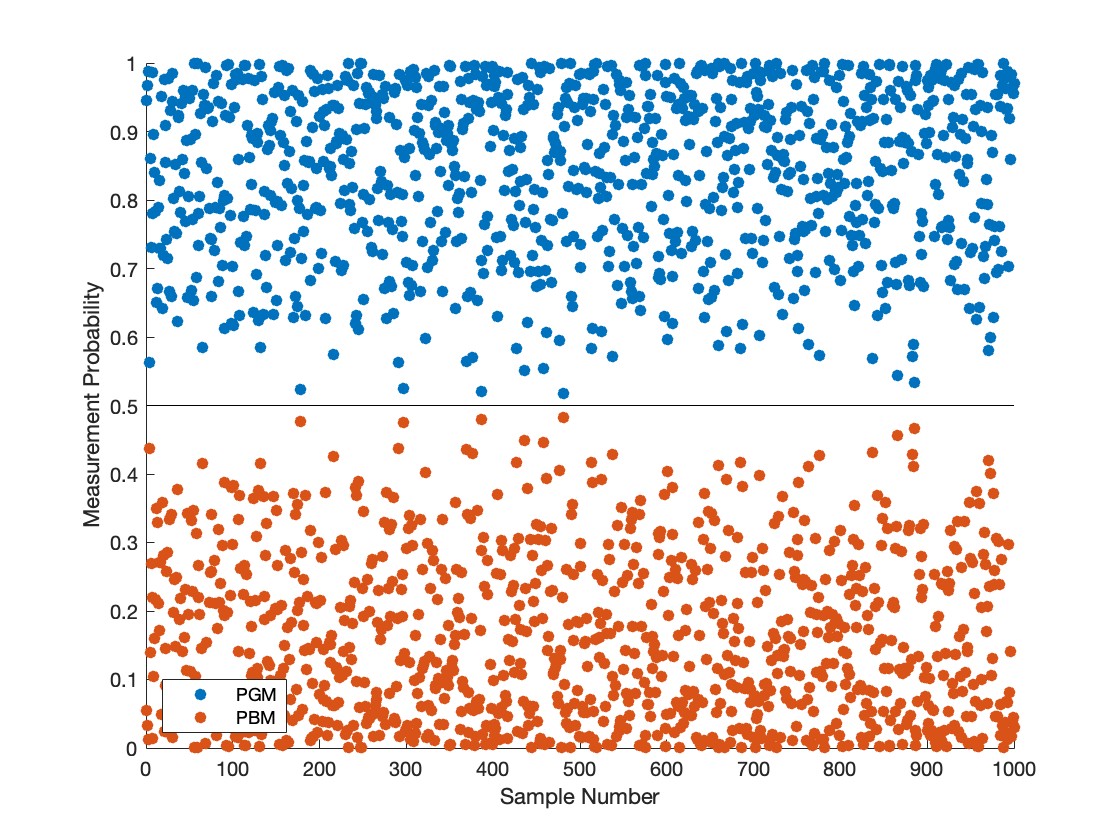}
    \caption{ 
    Computing PGM and PBM for $1000$ instances of $2$ randomly chosen qubits (also with randomly chosen probabilities).  
    Notice that PGM always performs better than $1/2$ and PBM always performs worse than $1/2$. 
    }
    \label{fig:pgm_vs_pbm_two_states}
\end{figure}

\begin{figure}[h]
    \centering 
\includegraphics[scale=0.220]{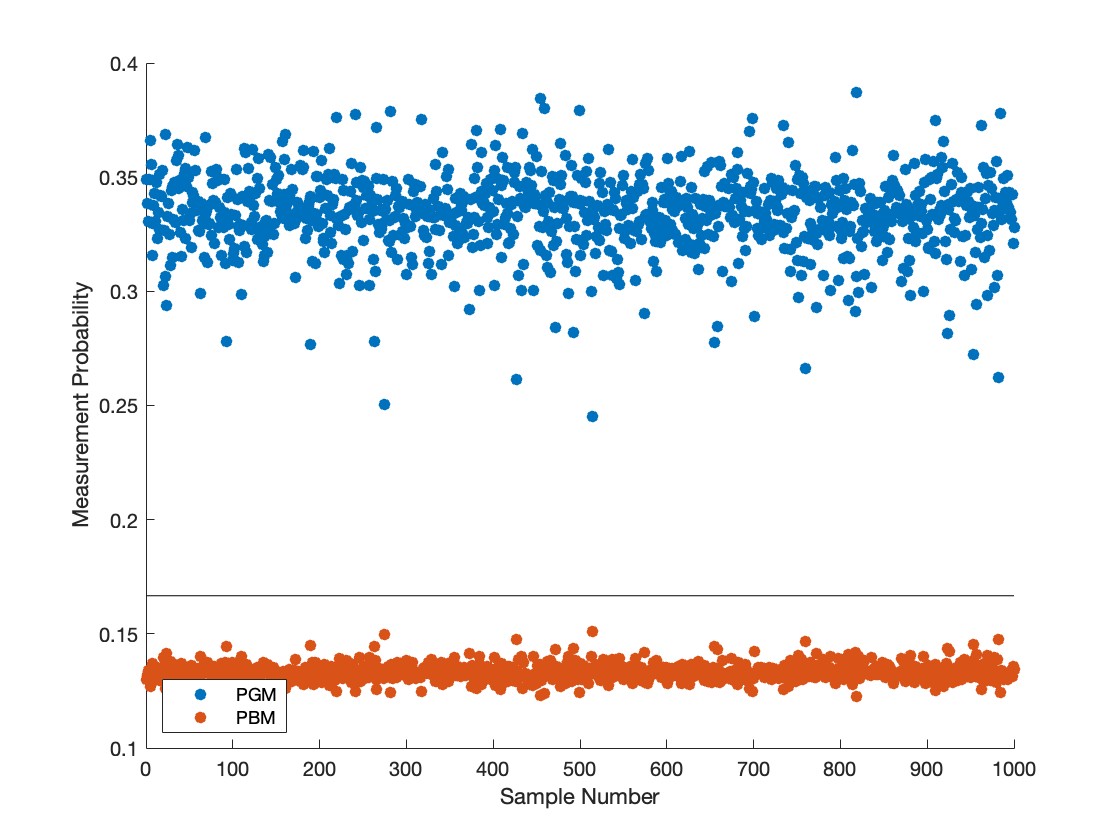}
    \caption{
    Computing PGM and PBM for $1000$ instances of $6$ randomly chosen qubits (also with randomly chosen probabilities).   
        We notice that $\PPGM$ seemingly clusters around $\frac{2}{k}$ and $\PPBM$ seemingly clusters around $\frac{k-2}{k(k-1)}$ for small values of $k$, with $k=6$ illustrated here. 
    } 
    \label{fig:pgm_vs_pbm_six_states}
\end{figure}

\medskip 
\paragraph{Numerical simulations.} 
We have discussed how the PGM always performs no worse than blind guessing and consequently proved that the PBM always performs no better than blind guessing.
To get a better idea of how these two probabilities compare, we calculate them on many randomly generated examples. 
To generate random states, we can use the Bloch sphere to define a parameterized qubit $\rho(\theta, \phi)$ as 
\begin{equation*}
    \rho(\theta, \phi) = \frac{1}{2} \mathds{1} + \frac{\sin(\theta)\cos(\phi)}{2} X + \frac{\sin(\theta)\sin(\phi)}{2}Y + \frac{\cos(\theta)}{2}Z 
\end{equation*}
where $X$, $Y$, and $Z$ are the Pauli operators and ${\theta, \phi \in (0, 2\pi)}$ are real parameters which we randomly generate. 
We generate $1000$ random qubit instances $\{ \rho_1, \ldots, \rho_k \}$ for $k=2$ in FIG.~\ref{fig:pgm_vs_pbm_two_states} and for $k=6$ in FIG.~\ref{fig:pgm_vs_pbm_six_states}, each with randomly generated probabilities $(p_1, \ldots, p_k)$. 

\medskip 
\paragraph{Application: Quantum anomaly detection/avoidance.} 
Suppose Alice promises to send Bob $k$ copies of a state $\rho$. 
However, she has a faulty device which is known to produce one copy of an anomaly (read: a \emph{pretty bad state}) $\sigma$ at some point in this sequence. 
Upon receiving the $k$ states, Bob wants to identify a state \emph{which is not faulty}, i.e., not $\sigma$. 

Our approach is to cast this as a discrimination problem with the states $\rho_i = \rho^{\otimes (i-1)}\ \otimes\ \sigma\ \otimes\ \rho^{\otimes (k-i)}$, for $i \in \{ 1, \ldots, k \}$, and to use the pretty bad measurement. 
Note that $\rho_i$ is simply $k$ copies of $\rho$, but with $\sigma$ in the $i$th spot. 
In the traditional discrimination task, Bob is trying to guess where the anomaly has occurred.
However, this is overkill if we just want to find a place where it \emph{did not occur}. 
Thus, taking the perspective of state exclusion makes more sense here. 
In other words, we do want to determine where the anomaly occurred as much as where it \emph{did not occur}. 
We focus on these two concepts in the following example and graphs. 

\begin{figure}[h]
    \centering
    \includegraphics[scale=0.575]{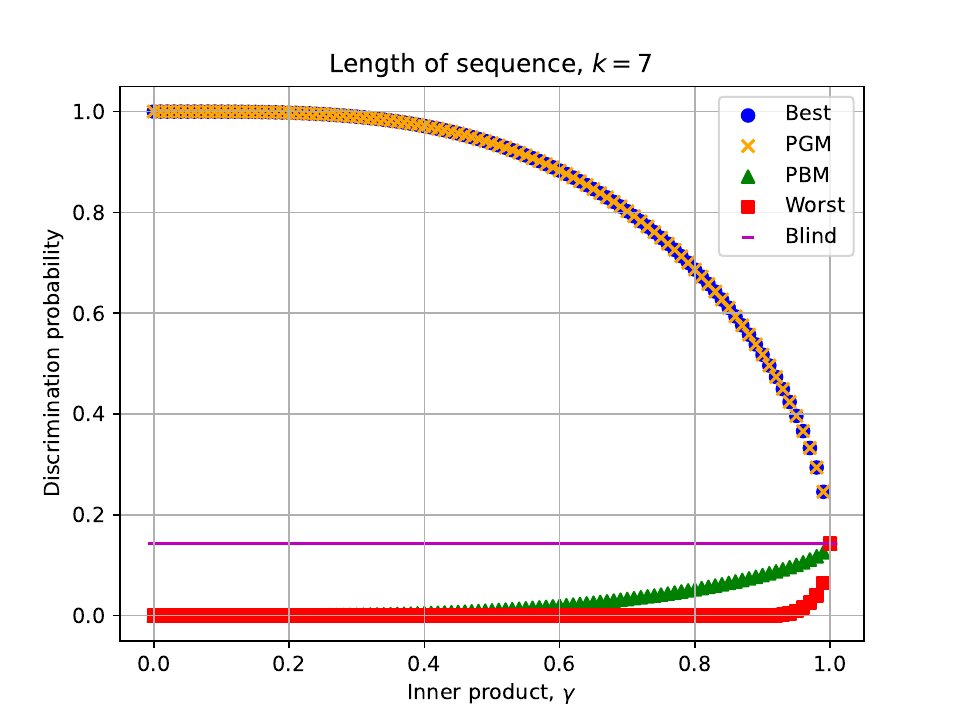}
    \caption{ 
    Probabilities for the quantum state discrimination task defined by the states  
    $\ket{\psi_i} = \ket{0}^{\otimes{i-1}} \otimes \ket{\psi} \otimes \ket{0}^{\otimes{7-i}}$, where $\ket{\psi} = \gamma \ket{0} + \sqrt{1 - \gamma^2} \ket{1}$ and $i \in \{ 1, \ldots, 7 \}$.  
    Observe that while $\PPGM = \Pbest$ for all values of $\gamma$, $\PPBM$ departs from the worst probability around $\gamma = 0.4$. 
    All of these probabilities coincide at $\gamma = 1$ since all the states are the same in this case (see Eq.~(\ref{allthesame})).  
    }
    \label{fig:anomaly_7}
\end{figure}

As an example, we consider the case where Alice promises to send to Bob $7$ copies of the state $\rho = \ketbra{0}{0}$ but an anomaly occurs creating the state $\sigma = \ketbra{\psi}{\psi}$ where $\ket{\psi} = \gamma \ket{0} + \sqrt{1 - \gamma^2} \ket{1}$ for some fixed parameter $\gamma \in [0,1]$ known to both Alice and Bob. 
FIG.~\ref{fig:anomaly_7} illustrates the probabilities $\Pbest$, $\PPGM$, $\PPBM$, $\Pworst$ and the blind guessing probability $1/k$, as a function of $\gamma$, below. 
For anomaly detection, \cite{llorens2023quantum, skotiniotis2018identification} showed that PGM is indeed the optimal measurement. Moreover they provide a closed-form expression for the optimal success probability. Therefore, using Eq.~\eqref{PGM_PBM} we also have an expression for $\PPBM$.

\begin{figure}[h]
    \centering
    \includegraphics[scale=0.575]{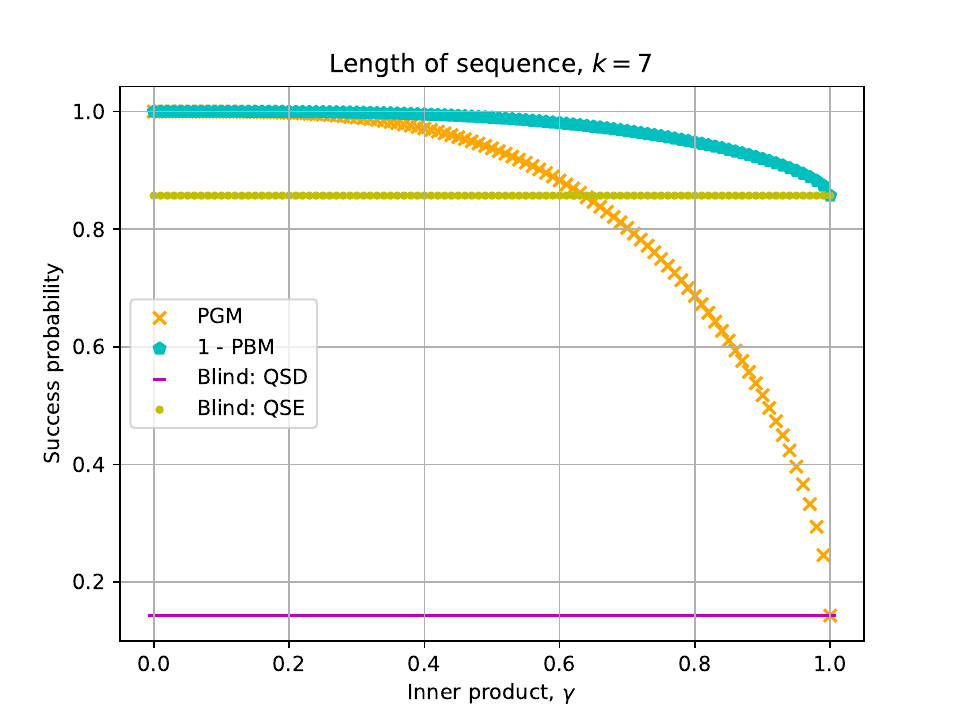}
    \caption{ 
    Given one of the states $\ket{\psi_i} = \ket{0}^{\otimes{i-1}} \otimes \ket{\psi} \otimes \ket{0}^{\otimes{7-i}}$, where $\ket{\psi} = \gamma \ket{0} + \sqrt{1 - \gamma^2} \ket{1}$, ranging over $i \in \{ 1, \ldots, 7 \}$, each with equal probability, we plot the probability of successfully avoiding the anomaly $\ket{\psi}$ via PBM and also the probability of successfully identifying $\ket{\psi}$ via PGM. 
    Notice that $\PPGM$ decreases rapidly starting around $\gamma = 0.2$, converging to the blind identification guessing  probability of $1/7$ when $\gamma = 1$. 
    On the other hand, $1 - \PPBM$ decreases at a much slower pace converging to the blind excluding probability of $6/7$ when $\gamma = 1$. 
    We remark that the exclusion probability is much larger than the identification probability, which is to be expected. 
    } 
    \label{fig:anomaly_7_success}
\end{figure}

Recall that in the anomaly avoidance problem, we want to identify a state that did not occur. 
Since PBM is trying to minimize $\sum_{i=1}^k \frac{1}{k} \tr(\rho_i M_i)$, what PBM is approximating in the context of anomaly avoidance is the minimum average error. 
In this case, the probability of \emph{success} using PBM is $1 - \PPBM$. 
In FIG.~\ref{fig:anomaly_7_success}, we compare this probability to that given by PGM which approximates the best average identification probability.  

\medskip 
\paragraph{Conclusions.} 
In this work we introduced the pretty bad measurement as a means to approximate the worst guessing probability for quantum state discrimination and compared it to the pretty good measurement. 
We showed that they have an intricate relationship and proved that each compare favorably to simply randomly guessing. 
Finally, we examined the quantum anomaly detection/avoidance problem and gave an illustrative example showing how these measurements perform for detecting/avoiding parameterized quantum states. 

\medskip 
\paragraph{Acknowledgements.} 
The authors thank Iman Marvian, Vincent Russo, Nathaniel Johnston, and Santiago Llorens for useful discussions. 
This research was funded in part by the Commonwealth of Virginia’s Commonwealth Cyber Initiative (CCI) under grant number~{467714}. 

\bibliographystyle{alpha}
\bibliography{PBM_arxiv_v2}

\newcommand{\etalchar}[1]{$^{#1}$}
\begin{thebibliography}{MCBK{\etalchar{+}}11}

\bibitem[ABGH02]{andersson2002minimum}
Erika Andersson, Stephen~M Barnett, Claire~R Gilson, and Kieran Hunter.
\newblock Minimum-error discrimination between three mirror-symmetric states.
\newblock {\em Physical Review A}, 65(5):052308, 2002.

\bibitem[Amb01]{ambainis2001new}
Andris Ambainis.
\newblock A new protocol and lower bounds for quantum coin flipping.
\newblock In {\em Proceedings of the thirty-third annual ACM symposium on
  Theory of computing}, pages 134--142, 2001.

\bibitem[AMNK21]{ahmed2021quantum}
Shahnawaz Ahmed, Carlos~S{\'a}nchez Mu{\~n}oz, Franco Nori, and Anton~Frisk
  Kockum.
\newblock Quantum state tomography with conditional generative adversarial
  networks.
\newblock {\em Physical Review Letters}, 127(14):140502, 2021.

\bibitem[BCvD06]{bacon2005optimal}
Dave Bacon, Andrew~M Childs, and Wim van Dam.
\newblock Optimal measurements for the dihedral hidden subgroup problem.
\newblock {\em Chicago Journal of Theoretical Computer Science}, (2), 2006.

\bibitem[Bel75a]{belavkin1975optimaldist}
Viacheslav~P Belavkin.
\newblock Optimal distinction of non-orthogonal quantum signals.
\newblock {\em Radio Eng. Electron. Phys.}, 20:39--47, 1975.

\bibitem[Bel75b]{belavkin1975optimal}
Viacheslav~P Belavkin.
\newblock Optimal multiple quantum statistical hypothesis testing.
\newblock {\em Stochastics: An International Journal of Probability and
  Stochastic Processes}, 1(1-4):315--345, 1975.

\bibitem[BJOP14]{bandyopadhyay2014conclusive}
Somshubhro Bandyopadhyay, Rahul Jain, Jonathan Oppenheim, and Christopher
  Perry.
\newblock Conclusive exclusion of quantum states.
\newblock {\em Physical Review A}, 89(2):022336, 2014.

\bibitem[BK15]{Bae_2015}
Joonwoo Bae and Leong-Chuan Kwek.
\newblock Quantum state discrimination and its applications.
\newblock {\em Journal of Physics A: Mathematical and Theoretical},
  48(8):083001, 2015.

\bibitem[Cab00]{cabello2000quantum}
Ad{\'a}n Cabello.
\newblock Quantum key distribution in the {H}olevo limit.
\newblock {\em Physical Review Letters}, 85(26):5635, 2000.

\bibitem[CPF{\etalchar{+}}10]{cramer2010efficient}
Marcus Cramer, Martin~B Plenio, Steven~T Flammia, Rolando Somma, David Gross,
  Stephen~D Bartlett, Olivier Landon-Cardinal, David Poulin, and Yi-Kai Liu.
\newblock Efficient quantum state tomography.
\newblock {\em Nature communications}, 1(1):149, 2010.

\bibitem[CR12]{christandl2012reliable}
Matthias Christandl and Renato Renner.
\newblock Reliable quantum state tomography.
\newblock {\em Physical Review Letters}, 109(12):120403, 2012.

\bibitem[{CVX}12]{cvx}
{CVX Research, Inc.}
\newblock {CVX}: Matlab software for disciplined convex programming, version
  2.0.
\newblock \url{http://cvxr.com/cvx}, August 2012.

\bibitem[DT10]{deconinck2010qubit}
Matthieu~E Deconinck and Barbara~M Terhal.
\newblock Qubit state discrimination.
\newblock {\em Physical Review A}, 81(6):062304, 2010.

\bibitem[EMV04]{eldar2004optimal}
Yonina~C Eldar, Alexandre Megretski, and George~C Verghese.
\newblock Optimal detection of symmetric mixed quantum states.
\newblock {\em IEEE Transactions on Information Theory}, 50(6):1198--1207,
  2004.

\bibitem[FHL17]{filippov2017necessary}
Sergey~N Filippov, Teiko Heinosaari, and Leevi Lepp{\"a}j{\"a}rvi.
\newblock Necessary condition for incompatibility of observables in general
  probabilistic theories.
\newblock {\em Physical Review A}, 95(3):032127, 2017.

\bibitem[GLF{\etalchar{+}}10]{gross2010quantum}
David Gross, Yi-Kai Liu, Steven~T Flammia, Stephen Becker, and Jens Eisert.
\newblock Quantum state tomography via compressed sensing.
\newblock {\em Physical Review Letters}, 105(15):150401, 2010.

\bibitem[HB20]{havlivcek2020simple}
Vojt{\v{e}}ch Havl{\'\i}{\v{c}}ek and Jonathan Barrett.
\newblock Simple communication complexity separation from quantum state
  antidistinguishability.
\newblock {\em Physical Review Research}, 2(1):013326, 2020.

\bibitem[Hel69]{helstrom1969quantum}
Carl~W Helstrom.
\newblock Quantum detection and estimation theory.
\newblock {\em Journal of Statistical Physics}, 1:231--252, 1969.

\bibitem[HJW93]{hughston1993complete}
Lane~P Hughston, Richard Jozsa, and William~K Wootters.
\newblock A complete classification of quantum ensembles having a given density
  matrix.
\newblock {\em Physics Letters A}, 183(1):14--18, 1993.

\bibitem[HKK06]{hayashi2006quantum}
Masahito Hayashi, Akinori Kawachi, and Hirotada Kobayashi.
\newblock Quantum measurements for hidden subgroup problems with optimal sample
  complexity.
\newblock {\em arXiv preprint quant-ph/0604174}, 2006.

\bibitem[JRS23]{johnston2023tight}
Nathaniel Johnston, Vincent Russo, and Jamie Sikora.
\newblock Tight bounds for antidistinguishability and circulant sets of pure
  quantum states.
\newblock {\em arXiv preprint arXiv:2311.17047}, 2023.

\bibitem[LD20]{leifer2020noncontextuality}
Matthew Leifer and Cristhiano Duarte.
\newblock Noncontextuality inequalities from antidistinguishability.
\newblock {\em Physical Review A}, 101(6):062113, 2020.

\bibitem[Led22]{leditzky2022optimality}
Felix Leditzky.
\newblock Optimality of the pretty good measurement for port-based
  teleportation.
\newblock {\em Letters in Mathematical Physics}, 112(5):98, 2022.

\bibitem[LSMT23]{llorens2023quantum}
Santiago Llorens, Gael Sentís, and Ramon Muñoz-Tapia.
\newblock Quantum multi-anomaly detection.
\newblock {\em arXiv preprint arXiv:2312.13020}, 2023.

\bibitem[MCBK{\etalchar{+}}11]{mallet2011quantum}
F~Mallet, MA~Castellanos-Beltran, HS~Ku, S~Glancy, E~Knill, KD~Irwin,
  GC~Hilton, LR~Vale, and KW~Lehnert.
\newblock Quantum state tomography of an itinerant squeezed microwave field.
\newblock {\em Physical Review Letters}, 106(22):220502, 2011.

\bibitem[MNW23]{mishra2023optimal}
Hemant~K Mishra, Michael Nussbaum, and Mark~M Wilde.
\newblock On the optimal error exponents for classical and quantum
  antidistinguishability.
\newblock {\em arXiv preprint arXiv:2309.03723}, 2023.

\bibitem[MR05]{moore2005distinguishing}
Cristopher Moore and Alexander Russell.
\newblock For distinguishing conjugate hidden subgroups, the pretty good
  measurement is as good as it gets.
\newblock {\em arXiv preprint quant-ph/0501177}, 2005.

\bibitem[MRD{\etalchar{+}}13]{mahler2013adaptive}
Dylan~H Mahler, Lee~A Rozema, Ardavan Darabi, Christopher Ferrie, Robin
  Blume-Kohout, and Aephraim~M Steinberg.
\newblock Adaptive quantum state tomography improves accuracy quadratically.
\newblock {\em Physical Review Letters}, 111(18):183601, 2013.

\bibitem[PBR12]{pusey2012reality}
Matthew~F Pusey, Jonathan Barrett, and Terry Rudolph.
\newblock On the reality of the quantum state.
\newblock {\em Nature Physics}, 8(6):475--478, 2012.

\bibitem[QHL{\etalchar{+}}13]{qi2013quantum}
Bo~Qi, Zhibo Hou, Li~Li, Daoyi Dong, Guoyong Xiang, and Guangcan Guo.
\newblock Quantum state tomography via linear regression estimation.
\newblock {\em Scientific Reports}, 3(1):3496, 2013.

\bibitem[Ren17]{renes2017better}
Joseph~M Renes.
\newblock Better bounds on optimal measurement and entanglement recovery, with
  applications to uncertainty and monogamy relations.
\newblock {\em Physical Review A}, 96(4):042328, 2017.

\bibitem[RJW{\etalchar{+}}08]{rippe2008experimental}
Lars Rippe, Brian Julsgaard, Andreas Walther, Yan Ying, and Stefan Kr{\"o}ll.
\newblock Experimental quantum-state tomography of a solid-state qubit.
\newblock {\em Physical Review A}, 77(2):022307, 2008.

\bibitem[RQK{\etalchar{+}}21]{rambach2021robust}
Markus Rambach, Mahdi Qaryan, Michael Kewming, Christopher Ferrie, Andrew~G
  White, and Jacquiline Romero.
\newblock Robust and efficient high-dimensional quantum state tomography.
\newblock {\em Physical Review Letters}, 126(10):100402, 2021.

\bibitem[RRS09]{radhakrishnan2009random}
Jaikumar Radhakrishnan, Martin R{\"o}tteler, and Pranab Sen.
\newblock Random measurement bases, quantum state distinction and applications
  to the hidden subgroup problem.
\newblock {\em Algorithmica}, 55:490--516, 2009.

\bibitem[RS23]{russo2023inner}
Vincent Russo and Jamie Sikora.
\newblock Inner products of pure states and their antidistinguishability.
\newblock {\em Physical Review A}, 107(3):L030202, 2023.

\bibitem[SBC{\etalchar{+}}16]{sentis2016quantum}
Gael Sent{\'\i}s, Emilio Bagan, John Calsamiglia, Giulio Chiribella, and Ramon
  Munoz-Tapia.
\newblock Quantum change point.
\newblock {\em Physical Review Letters}, 117(15):150502, 2016.

\bibitem[SCMT17]{sentis2017exact}
Gael Sent{\'\i}s, John Calsamiglia, and Ramon Munoz-Tapia.
\newblock Exact identification of a quantum change point.
\newblock {\em Physical Review Letters}, 119(14):140506, 2017.

\bibitem[SHCMT18]{skotiniotis2018identification}
M.~Skotiniotis, R.~Hotz, J.~Calsamiglia, and R.~Muñoz-Tapia.
\newblock Identification of malfunctioning quantum devices.
\newblock {\em arXiv preprint arXiv:1808.02729}, 2018.

\bibitem[Sik17]{sikora2017simple}
Jamie Sikora.
\newblock Simple, near-optimal quantum protocols for die-rolling.
\newblock {\em Cryptography}, 1(2):11, 2017.

\bibitem[SKR{\etalchar{+}}15]{schwemmer2015systematic}
Christian Schwemmer, Lukas Knips, Daniel Richart, Harald Weinfurter, Tobias
  Moroder, Matthias Kleinmann, and Otfried G{\"u}hne.
\newblock Systematic errors in current quantum state tomography tools.
\newblock {\em Physical Review Letters}, 114(8):080403, 2015.

\bibitem[SMP{\etalchar{+}}22]{stricker2022experimental}
Roman Stricker, Michael Meth, Lukas Postler, Claire Edmunds, Chris Ferrie,
  Rainer Blatt, Philipp Schindler, Thomas Monz, Richard Kueng, and Martin
  Ringbauer.
\newblock Experimental single-setting quantum state tomography.
\newblock {\em PRX Quantum}, 3(4):040310, 2022.

\bibitem[TMC{\etalchar{+}}18]{torlai2018neural}
Giacomo Torlai, Guglielmo Mazzola, Juan Carrasquilla, Matthias Troyer, Roger
  Melko, and Giuseppe Carleo.
\newblock Neural-network quantum state tomography.
\newblock {\em Nature Physics}, 14(5):447--450, 2018.

\bibitem[UBK{\etalchar{+}}20]{uola2020all}
Roope Uola, Tom Bullock, Tristan Kraft, Juha-Pekka Pellonp{\"a}{\"a}, and
  Nicolas Brunner.
\newblock All quantum resources provide an advantage in exclusion tasks.
\newblock {\em Physical Review Letters}, 125(11):110402, 2020.

\bibitem[VE02]{van2002unambiguous}
SJ~Van~Enk.
\newblock Unambiguous state discrimination of coherent states with linear
  optics: Application to quantum cryptography.
\newblock {\em Physical Review A}, 66(4):042313, 2002.

\end{thebibliography}

\appendix 

\section{Linear algebra background}  

We can view $\tr(XY)$ as an inner product on Hermitian matrices (called the Hilbert-Schmidt inner product) and the Frobenius norm is defined as the norm that this inner product induces, i.e., 
\begin{equation} 
\| X \|_F = \sqrt{\tr(X^2)}. 
\end{equation} 

The key component in the proof of our main result, Theorem 1, is the Cauchy-Schwarz inequality for Hermitian matrices, i.e., 
\begin{equation} 
|\tr(XY)| \leq \| X \|_F \| Y \|_F 
\end{equation} 
with equality if and only if $X$ and $Y$ are linearly dependent. 
Therefore, if \emph{nonzero} Hermitian $X$ and $Y$ satisfy 
\begin{equation} 
\tr(XY) = \| X \|_F \| Y \|_F 
\end{equation} 
then there must exist $\lambda \neq 0$ such that $X = \lambda Y$.

\section{Proof of a slightly weaker version of Theorem 1}  

Given the quantum states $\rho_1, \ldots, \rho_k$ and the probability distribution $(p_1, \ldots, p_k)$,  
define $P = \sum_{i=1}^k p_i \rho_i$ and the POVM $(G_1, \ldots, G_k)$ where $G_i = P^{-1/2} \left( p_i \rho_i \right) P^{-1/2}$ as in Eq.~\eqref{eq:PGM_ops}. 
Moreover, define the $k$-dimensional vector $r$ where 
\begin{equation} 
r_i = \| P^{-1/4} \left( p_i \rho_i \right) P^{-1/4}\|_F.  
\end{equation} 
The proof begins as follows. 
First, we establish that $\PPGM = \| r \|_2^2$. 
Second, we bound $\| r \|_1 \geq 1$. 
Then showing $\PPGM \geq 1/k$ follows by Cauchy-Schwarz.  

\bigskip 

To see that $\| r \|_2^2 = \PPGM$, we have 
\begin{align} 
\| r \|_2^2 
& = \sum_{i=1}^k r_i^2 \\ 
& = \sum_{i=1}^k \| P^{-1/4} \left( p_i \rho_i \right) P^{-1/4} \|_F^2 \\ 
& = \sum_{i=1}^k \tr \left( P^{-1/4} \left( p_i \rho_i \right) P^{-1/2} \left( p_i \rho_i \right) P^{-1/4} \right) \\ 
& = \sum_{i=1}^k p_i \ \tr \left( \rho_i \ P^{-1/2} \left( p_i \rho_i \right) \ P^{-1/2} \right) \\ 
& = \sum_{i=1}^k p_i \ \tr \left( \rho_i G_i \right) \\ 
& = \PPGM.  
\end{align} 

We now bound the following quantity:  
\begin{align} 
    & p_i \ \tr(\rho_i G_{i \oplus s}) \\ 
    & = p_i \ \tr(\rho_i \ P^{-1/2} (p_{i \oplus s} \rho_{i \oplus s})\ P^{-1/2} ) \\
    & = \tr\left( P^{-1/4} \left( p_i \rho_i \right) P^{-1/4} P^{-1/4} \left( p_{i \oplus s} \rho_{i \oplus s} \right) P^{-1/4} \right) \\ 
    & \leq \|  P^{-1/4} \left( p_i \rho_i \right) P^{-1/4} \|_F \| P^{-1/4} \left( p_{i \oplus s} \rho_{i \oplus s} \right) P^{-1/4} \|_F \label{temp} \\ 
    & = r_i \cdot r_{i \oplus s}. 
\end{align}

This bound is useful in showing $1 \leq \| r \|_1$. 
Using Eq.~\eqref{EqA3}, we have 
\begin{align}
    1 = \sum_{s=0}^{k-1} \alpha_s & = \sum_{s=0}^{k-1} \sum_{i=1}^k p_i \ \tr(\rho_i G_{i \oplus s}) \\
    & \leq \sum_{s=0}^{k-1} \sum_{i=1}^k r_i \cdot r_{i \oplus s} \\
    & = \left( \sum_{i=1}^k r_i \right)^2 \\ 
    & = \| r \|_1^2 
\end{align} 
noting the last equality holds since $r_i \geq 0$ for all $i$. 

\bigskip 

Let $e$ be the $k$-dimensional all-ones vector and note that $\| r \|_1 = \inner{r}{e}$. 
Then by the Cauchy-Schwarz inequality (for vectors), we have 
\begin{equation} \label{lastline}
1 \leq \| r \|_1 = \inner{r}{e} \leq \| r \|_2 \| e \|_2 = \sqrt{\PPGM} \cdot \sqrt{k} 
\end{equation} 
proving that $\PPGM \geq 1/k$.

\bigskip

Now, if we have that $\PPGM = 1/k$, we must have equality throughout~\eqref{lastline}. 
This means that the vector $r$ must be proportional to $e$ and $\| r \|_1 = 1$. 
These together imply that $r_i = 1/k$ for all $i$ which further implies that $P^{-1/4} (p_i \rho_i) P^{-1/4} \neq 0$ for all $i$, as well as $p_i \neq 0$ for all $i$. 

We further require inequality~\eqref{temp} to hold with equality, i.e., we must saturate Cauchy-Schwarz \emph{at every $i$ and $i \oplus s$}. 
This can only be true if there exists $\lambda_{i,j} \neq 0$ such that 
\begin{align}
    P^{-1/4} (p_i \rho_i) P^{-1/4} = \lambda_{i,j} P^{-1/4} (p_{j} \rho_{j}) P^{-1/4}   
\end{align} 
for all $i$ and $j$. 
Multiplying both sides by $P^{1/4}$, we have 
\begin{align} \label{thanos}
    p_i \rho_i = \lambda_{i,j} p_{j} \rho_{j}. 
\end{align} 
Taking the trace of both sides, we get 
\begin{align}
    \lambda_{i,j} = \frac{p_i}{p_j}. 
\end{align} 
From~\eqref{thanos}, we have $\rho_i = \rho_j$ for all $i$ and $j$. 

All that remains to prove is that $p_i = 1/k$  for all $i$. 
Recall that $r_i = 1/k$ for all $i$. 
Thus, we have 
\begin{align} 
1
& = \frac{r_i}{r_j} \\ 
& = \frac{\| P^{-1/4} (p_i \rho_i) P^{-1/4} \|_F}{\| P^{-1/4} (p_j \rho_j) P^{-1/4} \|_F} \\
& = \frac{p_i}{p_j} \frac{\| P^{-1/4} \rho_i P^{-1/4} \|_F}{\| P^{-1/4} \rho_j P^{-1/4} \|_F} \\ 
& = \frac{p_i}{p_j} 
\end{align}  
where the last equality holds since $\rho_i = \rho_j$ for all $i$ and $j$. 
Thus, $p_i = 1/k$ for all $i$, as required. 

\section{Proof of Lemma 1} 

Suppose $\PPBM = \Pworst$. 
Consider the \emph{best} measurement $(M_1, \ldots, M_k)$ and its random offset measurement probabilities $\alpha_s$. 
With this, we have $\alpha_0 = \Pbest$ and note that $\alpha_s \geq \Pworst = \PPBM$ for all $s$. 
By Eq.~\eqref{EqA3} we get
\begin{align}
1 
& = \sum_{s = 0}^{k-1} \alpha_s \\
& = \Pbest + \sum_{s = 1}^{k-1} \alpha_s \\ 
& \geq \Pbest + \sum_{s = 1}^{k-1} \PPBM \\ 
& \geq \PPGM + (k-1) \PPBM \\ 
& = 1  
\end{align}
where the last equality holds by Eq.~\eqref{PGM_PBM}. 
Thus, we have equality throughout, proving that $\PPGM = \Pbest$.   
  
\end{document}